%
%
\def\today{\ifcase\month\or January\or February\or March\or April\or May\or
June\or July\or August\or September\or October\or November\or December\fi
\space\number\day, \number\year}
%
%
\newcount\notenumber

\def\note{\global\advance\notenumber by 1 \footnote{$^{\the\notenumber}$}}
%
%
\newif\ifsectionnumbering
\newcount\eqnumber
\def\cleareqnumber{\eqnumber=0}
\def\numbereq{\global\advance\eqnumber by 1
\ifsectionnumbering\eqno(\the\secnumber.\the\eqnumber)
\else\eqno(\the\eqnumber) \fi}
\def\eqalinno{{\global\advance\eqnumber by 1}
\ifsectionnumbering(\the\secnumber.\the\eqnumber)\else(\the\eqnumber)\fi}
\def\name#1{\ifsectionnumbering\xdef#1{\the\secnumber.\the\eqnumber}
\else\xdef#1{\the\eqnumber}\fi}

\sectionnumberingtrue
%
%
\newcount\refnumber

\immediate\openout1=refs.tex
\immediate\write1{\noexpand\frenchspacing}
\immediate\write1{\parskip=0pt}
\def\ref#1#2{\global\advance\refnumber by 1%
[\the\refnumber]\xdef#1{\the\refnumber}%
\immediate\write1{\noexpand\item{[#1]}#2}}
%

%
%
\font\twelvebf=cmbx10 scaled \magstep1
\newcount\secnumber

\def\newsection#1.{\ifsectionnumbering\cleareqnumber\else\fi%
	\global\advance\secnumber by 1%
	\bigbreak\bigskip\par%
	\line{\twelvebf \the\secnumber. #1.\hfil}\nobreak\medskip\par}
%
%
%
\def \sqr#1#2{{\vcenter{\vbox{\hrule height.#2pt
	\hbox{\vrule width.#2pt height#1pt \kern#1pt
		\vrule width.#2pt}
		\hrule height.#2pt}}}}
\def\Box{{\mathchoice\sqr54\sqr54\sqr33\sqr23}\,}
%
%
%
\newdimen\fullhsize
\def\fiddle{\fullhsize=6.5truein \hsize=3.2truein}
\def\fullline{\hbox to\fullhsize}
\def\mkhdline{\vbox to 0pt{\vskip-22.5pt
	\fullline{\vbox to8.5pt{}\the\headline}\vss}\nointerlineskip}
\def\mkftline{\baselineskip=24pt\fullline{\the\footline}}
\let\lr=L \newbox\leftcolumn
\def\twocolumns{\fiddle
	\output={\if L\lr \global\setbox\leftcolumn=\columnbox
		\global\let\lr=R \else \doubleformat \global\let\lr=L\fi
		\ifnum\outputpenalty>-20000 \else\dosupereject\fi}}
\def\doubleformat{\shipout\vbox{\mkhdline
		\fullline{\box\leftcolumn\hfil\columnbox}
		\mkftline} \advancepageno}
\def\columnbox{\leftline{\pagebody}}
\magnification=1200
\def\pr#1 {Phys. Rev. {\bf D#1 }}
\def\pre#1 {Phys. Rep. {\bf #1}}
\def\pe#1 {Phys. Rev. {\bf #1}}
\def\pl#1 {Phys. Lett. {\bf B#1 }}
\def\prl#1 {Phys. Rev. Lett. {\bf #1 }}
\def\np#1 {Nucl. Phys. {\bf B#1 }}
\def\ap#1 {Ann. Phys. (NY) {\bf #1 }}
\def\cmp#1 {Commun. Math. Phys. {\bf #1 }}
\def\imp#1 {Int. Jour. Mod. Phys. {\bf A#1 }}
\def\mpl#1 {Mod. Phys. Lett. {\bf A#1}}
\def\BbbZ{{}\kern+1.6pt\hbox{$I$}\kern-7.5pt\hbox{$Z$}}

\def\s{(\sigma)}

\def\Tb{\overline T}

\def\ints{\int d\sigma\,}

\def\dx{\partial X}
\def\dbx{{\overline\partial}X}

\parskip=8pt plus 4pt minus 3pt
\headline{\ifnum \pageno>1\it\hfil  Higgs Mechanism in String Theory
	$\ldots$\else \hfil\fi}
\font\title=cmbx10 scaled\magstep1
\font\tit=cmti10 scaled\magstep1
\footline{\ifnum \pageno>1 \hfil \folio \hfil \else
\hfil\fi}
\raggedbottom
\rightline{\vbox{\hbox{RU96-1-B}\hbox{JHU-TIPAC-97005}}}
\vfill
\centerline{\title HIGGS MECHANISM IN STRING THEORY}
\vfill
{\centerline{\title Jonathan Bagger${}^{a}$
and
Ioannis Giannakis${}^{b}$ \footnote{$^{\dag}$}
{\rm e-mail: bagger@jhu.edu,
giannak@theory.rockefeller.edu}}
}
\medskip
\centerline{$^{(a)}${\tit Department of  Physics and Astronomy}}
\centerline{\tit The Johns Hopkins University}
\centerline{\tit Baltimore, MD 21218}
\medskip
\centerline{$^{(b)}${\tit Physics Department, Rockefeller
University}}
\centerline{\tit 1230 York Avenue, New York, NY
10021-6399.}
\vfill
\centerline{\title Abstract}
\bigskip
{\narrower\narrower
In first-quantized string theory, spacetime symmetries are described
by inner automorphisms of the underlying conformal field theory.  In
this paper we use this approach to illustrate the Higgs effect in string
theory.  We consider string propagation on $M^{24,1} \times S^1$, where
the circle has radius $R$, and study $SU(2)$ symmetry breaking as $R$
moves away from its critical value.  We find a gauge-covariant equation
of motion for the broken-symmetry gauge bosons and the would-be Goldstone
bosons.  We show that the Goldstone bosons can be eliminated by an
appropriate gauge transformation.  In this unitary gauge, the Goldstone
bosons become the longitudinal components of massive gauge bosons.
\par}
\vfill\vfill\break

\newsection Introduction.

String theory remains the most promising candidate for a unified
description of nature.  During the past few years, many string
dualities have been discovered, but it is fair to say that a deep
understanding of string dynamics is still lacking.  It is therefore
important to understand the role of spacetime symmetries in string
dynamics.

At the classical level, every two-dimensional conformal field theory
(of the appropriate central charge) is a solution to the string
equations of motion
\ref\cala{C. Lovelace, \pl135\ (1984) 75; C. Callan, D. Friedan,
E. Martinec and M. Perry, \np262\ (1985) 593; A. Sen, \pr32\ (1985)
2102; A. Belavin, A. Polyakov and A. Zamolodchikov, \np241\ (1984)
333; D. Friedan, E. Martinec and S. Shenker, \np271\ (1986) 93.}.
The two-dimensional couplings are the spacetime fields of the string.
Conformal invariance determines the couplings, hence the dynamics of
the spacetime fields.  During the past decade, a large number of
string solutions have been constructed in this way
\ref\sol{P. Candelas, G. Horowitz, A. Strominger and E. Witten,
\np258\ (1985) 46; L. Dixon, J. Harvey, C. Vafa and E. Witten,
\np261\ (1985) 678; \np274\ (1986) 285; K. Narain, \pl169\ (1986)
41; H. Kawai, D. Lewellen and S. Tye, \np288\ (1987) 1;
I. Antoniadis, C. Bachas and C. Kounnas, \np289\ (1987) 87;
I. Antoniadis and C. Bachas, \np298\ (1988) 586; K. Narain,
H. Sarmadi and C. Vafa, \np288\ (1987) 551; R. Bluhm, L. Dolan
and P. Goddard, \np289\ (1987) 364; W. Lerche, A. Schellekens and
N. Warner, \pre177\ (1989) 1.}.

Fortunately, many of the string solutions are related by symmetries.
In ordinary field theory, symmetries are transformations of the
spacetime fields which leave the classical action invariant.  Barring
anomalies, they also hold in the full quantum theory.  In string theory,
the situation is different.  The spacetime fields appear as couplings,
so symmetries are not invariances of a spacetime action.

There are good reasons to believe that string theory contains an
enormous degree of symmetry, of which gauge and coordinate invariance
are but remnants.  First, the particle content and interactions
of string theory are so tightly constrained that they are presumably
fixed by some symmetry.  Second, high-energy fixed-angle string
scattering obeys a universal behavior which suggests that some
large symmetry is being restored
\ref\gross{D. Gross, \prl60\ (1988) 1229; D. Gross and
P. Mende, \np303\ (1988) 407; D. Gross and J. Manes,
\np326\ (1989) 73.}.
This symmetry mixes massless and massive states, and is
spontaneously broken by the vacuum.  Other aspects of
symmetry breaking in string theory are discussed in
\ref\ant{I. Antoniadis, C. Bachas and C. Kounnas, \pl200
\ (1988) 297; S. Ferrara, C. Kounnas and M. Porrati, \pl206
\ (1988) 25; L. Ibanez, W. Lerche, D. Lust and S. Theisen,
\np352\ (1991) 435; M. Porrati, \pl231\ (1989) 403;
G. Moore, "Symmetries and Symmetry Breaking in
String Theory," hep-th/9308052;
E. Kiritsis, \np405\ (1993) 109.}.

Recently, a simple but powerful approach to string symmetries
was developed in
\ref{\evao}{M. Evans and B. Ovrut \pr41\ (1990) 3149;
\pl231\ (1989) 80.}.
In this work, string symmetries are identified with similarity
transformations of the underlying conformal field theory.
The key idea is that automorphisms of the operator algebra
change the Hamiltonian, but do not affect the physical
results.

The approach of Ref.~[\evao] is very general.  It gives rise to
spacetime symmetries which mix states of different mass.  It
places unbroken and spontaneously broken spacetime symmetries
on exactly the same footing.  Unbroken spacetime symmetries are
generated by conserved currents of the underlying conformal
field theory
\ref{\band}{T. Banks and L. Dixon \np307\ (1988) 93.},
while spontaneously broken symmetries are generated by currents
that are not conserved.

In this paper we will study spontaneously broken symmetries in
string theory.  We will focus on a simple example: string
propagation on $M^{24, 1} \times S^1$.  For a generic value of
the radius of $S^1$, this string vacuum has an unbroken $U(1)_L
\times U(1)_R$ gauge symmetry.  At a critical value of $R$, the
symmetry is enhanced to $SU(2)_L \times SU(2)_R$.  Away from
this critical value, the $SU(2)_L \times SU(2)_R$ is
spontaneously broken to $U(1)_L \times U(1)_R$.

In field theory, the spontaneous breaking of a gauge symmetry
is associated with the Higgs effect, through which the would-be
Goldstone bosons are absorbed by the spontaneously-broken gauge
bosons.  The physical spectrum is manifest in unitary gauge,
where one finds a set of massive gauge bosons, one for each
spontaneously broken generator.

The formalism of [\evao] is especially well-suited for describing
the Higgs effect in string theory.  Therefore, in what follows, we
will first review the status of symmetries in (perturbative) string
theory.  We will then restrict our attention to string propagation
on $M^{24, 1} \times S^1$.  We will start at the critical radius
and identify the generators of the $SU(2)_L \times SU(2)_R$
gauge symmetry.  We will find the full set of massless scalar fields,
as well as the massless gauge bosons associated with the unbroken
gauge symmetry.  We will then shift away from the critical radius
by giving a small expectation value to a modulus field.  We will
see that this vev spontaneously breaks the $SU(2)_L \times SU(2)_R$
symmetry to $U(1)_L \times U(1)_R$.  We will find that the scalar
multiplets split into Goldstone and physical fields, and discover
how the gauge bosons absorb the Goldstone modes.  We shall see
that this model illustrates one particularly simple way in which
the full set of string symmetries is broken by the string vacuum.

\bigbreak\bigskip

\newsection Symmetries in (Perturbative) String Theory.

In ordinary string theory, the classical string solutions are in
one-one correspondence with conformal field theories of the appropriate
central charge.  Given one string solution, a physically equivalent
solution can be found by making a similarity transformation on the
operator algebra $\cal A$ of the conformal field theory [\evao],
$$
\Phi(\sigma) \mapsto
e^{ih}\Phi(\sigma)e^{-ih}.
\numbereq\name{\eqinaut}
$$
This determines an equivalent solution for any operator $h$.

This automorphism (\eqinaut) acts on the stress tensor in the
obvious way,
$$
T_{\phi}(\sigma)\mapsto
e^{ih}T_{\phi}(\sigma)e^{-ih}
\numbereq\name{\eqinopq}
$$
(and likewise for $\Tb_\phi\s$), where $\phi$ denotes a generic
spacetime field.  In what follows, we shall restrict our attention
to automorphisms which change the spacetime fields.  Therefore, we
require
$$T_{\phi+\delta \phi}\s-T_\phi\s= i [h,T_\phi\s],
\numbereq\name{\eqh}
$$
for some infinitesimal operator $h$.  The transformation $\phi
\mapsto \phi+ \delta\phi$ is a {\it symmetry}:  it is an
infinitesimal change of the spacetime fields which does not
change the physics.

{}From this point of view, symmetries are infinitesimal
deformations of the stress tensor, $T_\phi\s \mapsto T_\phi\s +
\delta T$, where $\delta T = i [h,T_\phi\s]$.  More general
deformations are not symmetry transformations, but describe
physically distinct solutions.  For example, two nearby solutions
are flat space-time $M^{24,1} \times S^1$, and a weak electromagnetic
wave propagating through it.  These two solutions are not related
by any symmetry transformation.

For string propagation on $M^{24, 1} \times S^1$, the vacuum stress
energy tensor is given by
$$
T(\sigma)=-{1\over 2} \eta_{\mu\nu} \partial X^\mu \partial X^\nu -
{1\over 2} \partial X^{26} \partial X^{26}
\numbereq\name{\free}
$$
at the radius $R = R_{cr} = \sqrt 2$.
A weak $U(1)_L$ electromagnetic wave can be obtained by adding
$$
\delta T =-A^{(3)}_\mu(X)\dbx^\mu\dx^{26}.
\numbereq\name{\eqd}
$$
This deformation preserves conformal invariance provided
$\delta T$ is a primary field of dimension $(1,1)$.  This is equivalent
to saying that the functions $A^{(3)}_\mu(X)$ satisfy the following
conditions,
$$
\Box A^{(3)}_\mu(X)=0\qquad
\partial^\mu A^{(3)}_\mu(X)=0.
\numbereq\name{\eqgauge}
$$
The first is an equation of motion; the second is a Lorentz
gauge condition.

Equations (\eqd) and (\eqgauge) provide an example of a {\it
canonical deformation} [\evao], defined by $\delta T\s=V(\sigma)$,
where $V(\sigma)$ is a vertex operator,  a primary field
of dimension $(1,1)$.  Canonical deformations take one conformal
field theory into another.  Furthermore, they induce a variation
in the stress tensor which can be expressed as a change in the
spacetime fields.

Canonical deformations turn on gauge fields in the Lorentz gauge.
To describe the Higgs effect, however, we would like to
transform the spacetime fields to an {\it arbitrary} gauge
\ref\gvan{M. Evans and I. Giannakis \pr44\ (1991) 2467;
\np472\ (1996) 139; I. Giannakis, \pl388\ (1996) 543;
M. Evans, I. Giannakis and D. V. Nanopoulos, \pr50
\ (1994) 4022.}.
This can be achieved by performing an automorphism $\delta T =
i[h,T_\phi\s]$, where $h$ is given by
$$
h=\ints \Lambda^{(3)}(X) {i \sqrt{2} \partial X^{26}}
\numbereq\name{\eqasovc}
$$
and ${\Lambda^{(3)}}(X)$ is the parameter of the gauge
transformation.  Note that if $\Box{\Lambda^{(3)}}=0$, the
integrand is of dimension $(1, 0)$, and the transformation
(\eqasovc) preserves the Lorentz gauge.

To see this, let us start with the field $A^{(3)}_\mu(X)$ in Lorentz
gauge,
$$
T(\sigma)=-{1\over 2} \eta_{\mu\nu} \partial X^\mu \partial X^\nu -
{1\over 2} \partial X^{26} \partial X^{26} -
A^{(3)}_\mu(X) \overline{\partial} X^\mu \partial X^{26},
\numbereq\name{\eqwianc}
$$
where $\Box A^{(3)}_\mu(X) = \partial^\mu A^{(3)}_\mu(X)=0$.
Let us then compute the commutator $i[h, T(\sigma)]$.  This gives
rise to the following deformed stress energy tensor,
$$
\eqalign{
T'\s  =&\
T(\sigma)+i[h, T(\sigma)]\cr
=& -{1\over 2}\eta_{\mu\nu}
\partial X^\mu \partial X^{\nu}-
{1\over 2}{\partial}X^{26}{\partial}X^{26} \cr
& -(A^{(3)}_\mu(X)+{\partial_\mu}{\Lambda^{(3)}}(X))
{\overline{\partial}}X^\mu{\partial}X^{26}
-{1\over 2}\Box\Lambda^{(3)}(X){\partial^2}X^{26} \cr
& -{1\over 2}\Box{\partial_\mu}
\Lambda^{(3)}(X){\partial}X^\mu{\partial}X^{26}+
{1\over 2}\Box{\partial_\mu}
\Lambda^{(3)}(X){\overline{\partial}}X^\mu{\partial}X^{26}.}
\numbereq\name{\eqduios}
$$
Writing
$$
A'^{(3)}_\mu(X) = A^{(3)}_\mu(X) + \partial_\mu\Lambda^{(3)}(X)
\numbereq\name{\gaugetr}
$$
and imposing the conformal condition (\eqgauge), we find the
general stress tensor
$$
\eqalign{
T'\s =& -{1\over 2}\eta_{\mu\nu}
\partial X^\mu \partial X^{\nu}-
{1\over 2}{\partial}X^{26}{\partial}X^{26} \cr
& - A'^{(3)}_\mu(X){\overline{\partial}}X^\mu{\partial}
X^{26} - {1\over 2}{\partial^\mu}A'^{(3)}_\mu(X){\partial^2}X^{26}\cr
& - {1\over 2}\Box A'^{(3)}_\mu(X){\partial}X^\mu{\partial}X^{26}
+ {1\over 2}\Box A'^{(3)}_\mu(X){\overline{\partial}}X^\mu{\partial}
X^{26}.}
\numbereq\name{\eqohfk}
$$
and the gauge-covariant equations of motion
$$
\Box A'^{(3)}_\mu(X)-{\partial_\mu}{\partial^{\nu}}
A'^{(3)}_{\nu}(X)=0.
\numbereq\name{\gieom}
$$
Note that this gauge-invariant equation of motion reduces to (\eqgauge)
when $\partial^\mu A'^{(3)}_\mu(X) = 0$.

\bigbreak\bigskip
\newsection Strings on $M^{24,1} \times S^1$.

In this section we will take a closer look at bosonic string propagation
on $M^{24,1} \times S^1$.  We take $X^{26}$ to be periodic: $X^{26} \sim
X^{26} + 2{\pi}R$, where $R$ is the radius of the circle $S^1$.  On this
space there are two types of excitations: strings with quantized momenta
in the compact dimension, and strings that wind around the compact dimension
a fixed number of times.  The mass formula for the $25$-dimensional
particle states receives contributions from both,
$$
M^2={n^2 \over R^2}+{{m^2 R^2}\over 4}+N_L+N_R-2,
\numbereq\name{\eqsurew}
$$
where $m$ and $n$ are integers, and $N_L (N_R)$ denote the oscillator
contributions from the left (right) sectors.  Physical string states must
also satisfy the reparametrization constraint $N_L-N_R=mn$.

The space $M^{24,1} \times S^1$ is a consistent string vacuum for arbitrary
radius $R$.  The vacuum stress tensor is
$$
T_R(\sigma)=-{1\over 2} \eta_{\mu\nu} \hat{\partial} X^\mu
\hat{\partial} X^\nu -
{1\over 2}{R^2\over R_{cr}^2} \hat{\partial} X^{26}
\hat{\partial} X^{26},
\numbereq\name{\eqfuxo}
$$
together with its conjugate $\Tb_R\s$.  The operator $\hat{\partial}$
is the usual light-cone derivative.  Note that $\hat{\partial} X^\mu$
remains invariant as the radius $R$ is varied, but that the operator
$\hat{\partial} X^{26}$ does not.  This is easy to see by writing the
operators in terms of the string coordinates $(X^{\mu}, X^{26})$,
together with the conjugate momenta $(\pi_{\mu},\pi_{26})$,
$$
\eqalign{
\hat{\partial} X^\mu&={1\over {\sqrt 2}}(\eta^{\mu\nu}
\pi_\nu + X'^\mu)\cr
\hat{\overline{\partial}}X^\mu&=
{1\over {\sqrt 2}}(\eta^{\mu\nu}\pi_\nu - X'^\mu)\cr}
\qquad \eqalign{
\hat{\partial} X^{26}&={1\over {\sqrt 2}}\left({R_{cr}^2\over
R^2}\pi_{26} +X'^{26}\right)\cr
\hat{\overline{\partial}}X^{26}&={1\over {\sqrt 2}}
\left({R_{cr}^2\over R^2}\pi_{26}- X'^{26}\right). \cr}
\numbereq\name{\eqsmz}
$$

To exhibit the Higgs effect, we will need to compare conformal
field theories at different radii -- that is, at different values of
the background fields.  It is therefore essential to express the
stress tensors in terms of fixed, background-independent operators,
such as $\pi_{26}$ and $X^{26}$.  Therefore, in what follows, we {\it
define} the symbols $\partial X^\mu$, $\partial X^{26}$, $\overline
{\partial} X^\mu$ and $\overline{\partial}X^{26}$ (without the hats)
to be
$$
\eqalign{
\partial X^\mu &={1\over {\sqrt 2}}(\eta^{\mu\nu}
\pi_\nu + X'^\mu)\cr
{\overline{\partial}}X^\mu &=
{1\over {\sqrt 2}}(\eta^{\mu\nu}\pi_\nu - X'^\mu)\cr}
\qquad \eqalign{
\partial X^{26}&={1\over {\sqrt 2}}(\pi_{26}
+X'^{26})\cr
{\overline{\partial}} X^{26}&={1\over {\sqrt 2}}
(\pi_{26}- X'^{26}). \cr}
\numbereq\name{\eqdxoi}
$$
These operators have fixed, radius-independent commutation relations.
At the critical radius, $R=R_{cr}={\sqrt 2}$, they reduce to the
light-cone derivatives.

The light-cone derivatives $\hat{\partial} X^\mu$ and $\hat{\partial}
X^{26}$ can be expressed in terms of the fixed operators (\eqdxoi).
The result is
$$
\hat{\partial} X^\mu=\partial X^\mu, \qquad
\hat{\partial} X^{26}={1\over 2}\left[
{{R_{cr}^2}\over {R^2}}({\partial X}^{26}+{\overline{\partial}}X^{26})+
(\partial X^{26}
-{\overline{\partial}}X^{26})\right].
\numbereq\name{\eqduax}
$$
When substituted into eq.~(\eqfuxo), they give the stress tensor in
the fixed basis,
$$
\eqalign{
T_R=&-{1\over 2}\eta_{\mu\nu} \partial X^\mu \partial X^\nu
-{1\over 8} \bigg[ \left({R_{cr}\over R}+{R\over R_{cr}}\right)^2
\partial X^{26}\partial X^{26}\cr
& -\left({R_{cr}\over R}-{R\over R_{cr}}\right)^2
\overline{\partial}X^{26}\overline{\partial}X^{26}
-2\left({{R_{cr}}^2\over R^2}-{R^2\over {R_{cr}}^2}\right)
\partial X^{26}\overline{\partial}X^{26} \bigg]. \cr}
\numbereq
$$
For small variations $R = R_{cr} + \delta R$, one finds
$$
\eqalign{
\hat{\partial} X^{26}=&\partial X^{26}
\left(1-{{\delta}R\over R_{cr}}
+3{({\delta}R)^2\over 2R_{cr}^2}\right)
+{\overline{\partial}}X^{26}\left(-{{\delta}R\over R_{cr}}
+3{({\delta}R)^2\over 2R_{cr}^2}\right)\cr
\hat{\overline{\partial}}X^{26}=&\partial X^{26}
\left(-{{\delta}R\over R_{cr}}
+3{({\delta}R)^2\over 2R_{cr}^2}\right)+\overline{\partial}X^{26}
\left(1-{{\delta}R\over R_{cr}}
+3{({\delta}R)^2\over 2R_{cr}^2}\right)\cr}
\numbereq\name{\eqamopzy}
$$
and
$$
\eqalign{
T_R=&-{1\over 2} \eta_{\mu\nu} \partial X^\mu \partial X^\nu -
\left({1\over 2}+{({\delta}R)^2\over 2R_{cr}^2}\right)
\partial X^{26} \partial X^{26}\cr
& -{({\delta}R)^2\over 2R_{cr}^2}
{\overline{\partial}}X^{26} {\overline\partial}X^{26}
-\left(-{{\delta}R\over R_{cr}}+{({\delta}R)^2\over 2R_{cr}^2}\right)
{\partial}X^{26}{\overline{\partial}}X^{26}. \cr}
\numbereq\name{\eqsciopp}
$$

The vertex operators for the emission or absorption of particle
states should also be written in terms of the fixed basis.  At
the critical radius, the vertex operators
$$
\eqalign{
V^{(3)}(\sigma)=&A^{(3)}_\mu(X){\overline{\partial}}X^\mu
\partial X^{26}  \cr
\tilde{V}^{(3)}(\sigma)
=&\tilde{A}^{(3)}_\mu(X) \partial X^\mu{\overline{\partial}}X^{26}\cr
V^{(33)}(\sigma)=&\phi^{(33)}(X)
\partial X^{26}{\overline{\partial}}X^{26} }
\numbereq\name{\eqaposqz}
$$
are subject to the conditions
$$
\Box A^{(3)}_\mu(X)=\Box \tilde{A}^{(3)}_\mu(X)=
\Box  \phi^{(33)}(X)=0 \qquad \partial^{\nu}
A^{(3)}_{\nu}(X)=\partial^{\nu}\tilde{A}^{(3)}_{\nu}(X)
=0.
\numbereq\name{\primary}
$$
They describe the emission or absorption of massless gauge and scalar
bosons.  As in sect.~2, the vertex operators create gauge bosons in
Lorentz gauge.

For $R = R_{cr}$, there are other massless particles in the spectrum.
They include four massless gauge bosons, whose vertex operators are
given by
$$
{V^{(\pm)}}(\sigma)= A^{(\pm)}_\mu(X) \overline{\partial}X^\mu
\exp({\pm i}{\sqrt 2}X^{26}_L) \qquad \tilde{V}^{(\pm)}(\sigma)=
{\tilde A}^{(\pm)}_\mu(X) \partial X^\mu
\exp({\pm i}{\sqrt 2}X^{26}_R),
\numbereq\name{\eqazuiew}
$$
together with eight massless scalars, whose vertex operators take
the form
$$
\eqalign{
V^{(3\pm)}(\sigma)=& \phi^{(3\pm)}(X)\partial X^{26}
\exp({\pm i}{\sqrt 2}X^{26}_R) \cr
\tilde V^{(\pm 3)}(\sigma)=&
\phi^{(\pm 3)}(X){\overline{\partial}}X^{26}\exp({\pm i}{\sqrt 2}X^{26}_L) \cr
V^{(\pm\pm)}(\sigma)=&\phi^{(\pm\pm)}(X)
\exp({\pm i}{\sqrt 2}X^{26}_L)\exp({\pm i}{\sqrt 2}X^{26}_R)  \cr
V^{(\pm\mp)}(\sigma)=&\phi^{(\pm\mp)}(X)
\exp({\pm i}{\sqrt 2}X^{26}_L)\exp({\mp i}{\sqrt 2}X^{26}_R).  \cr}
\numbereq\name{\eqzopiua}
$$
The fields $A^{(\alpha)}_\mu(X)$, $\tilde A^{(\alpha)}_\mu(X)$ and
${\phi}^{(\alpha\beta)}(X)$ satisfy the conformal conditions
$$
\Box A^{(\alpha)}_\mu(X)=\Box {\tilde A}^{(\alpha)}_\mu(X)=
\Box  {\phi}^{(\alpha\beta)}(X)=0 \qquad {\partial^{\nu}}
A^{(\alpha)}_{\nu}(X)={\partial^{\nu}}{\tilde A}^{(\alpha)}_{\nu}(X)
=0,
\numbereq\name{\eqsuziop}
$$
for $\alpha,\beta = 3,\pm$.  As above, they are massless equations
of motion and Lorentz gauge conditions for the gauge and scalar
bosons.  The full set of massless gauge bosons fills out the
adjoint representation of $SU(2)_L \times SU(2)_R$; the massless
scalars transform in the $(3,3)$ representation of the gauge group.

If one deforms the conformal field theory by varying the radius
of the circle, the vertex operators change continuously.
The new vertex operators are as above, with $\partial X^{26}$
and $\overline{\partial} X^{26}$ replaced by the operators
$\hat{\partial} X^{26}$ and $\hat{\overline{\partial}} X^{26}$.
(The operators $\partial X^\mu$ and $\overline{\partial} X^\mu$
do not depend on $R$.)

For arbitrary radius $R$, the
condition that the deformed vertex operators be $(1,1)$ primary
fields with respect the deformed stress tensor, eq.~(\eqsciopp),
gives rise to massless equations of motion and Lorentz gauge
conditions for the following spacetime fields,
$$
\Box A^{(3)}_\mu(X)=\Box \tilde A^{(3)}_\mu(X)=
\Box \phi^{(33)}(X)=0 \qquad \partial^{\nu}
A^{(3)}_{\nu}(X)=\partial^\nu \tilde A^{(3)}_{\nu}(X)
=0.
\numbereq\name{\eqsnxoiop}
$$
In contrast, $A^{(\pm)}_\mu(X)$, $\tilde A^{(\pm)}_\mu(X)$,
$\phi^{(3\pm)}(X)$, and $\phi^{(\pm 3)}(X)$ obey massive equations of
motion and modified, $R_\xi$-like gauge conditions
$$
\eqalign{
&\Box A^{(\pm)}_\mu(X)+{(\delta R)^2\over 2}A^{(\pm)}_\mu(X)
=0 \qquad \Box \phi^{(\pm3)}(X)+
{(\delta R)^2\over 2}{\phi}^{(\pm3)}(X)=0 \cr
&\partial^{\nu}A^{(\pm)}_{\nu}(X)=-
\left({\delta R\over R_{cr}^2}-{(\delta R)^2\over 2R_{cr}^3}\right)
{\phi}^{(\pm3)}(X). \cr}
\numbereq\name{\eqaxoin}
$$

\newsection Higgs Mechanism in String Theory.

We are now ready to exhibit the string theory Higgs effect.  We
first need to relax the $R_\xi$-like gauge condition.  As in the
previous section, we can do this by carrying out a general
$SU(2)_L$ (or $SU(2)_R$) gauge transformation on the spacetime
fields.

The $SU(2)_L$ gauge transformation is most easily specified at
the critical radius.  It is generated by an operator $h$,
$$
h = \ints \Lambda^{(a)}(X) J^{(a)}(\sigma),
\numbereq
$$
where the dimension $(1,0)$ currents $J^{(a)}$, $a = 1,2,3,$ are
conserved.  For the case at hand, the $SU(2)_L$ currents are simply
$$
J^{(3)}\s =  i{\sqrt 2}{\partial}X, \quad
J^{(\pm)}\s = \exp(\pm i \sqrt 2 X_L).
$$
Therefore the operator $h$ can be written as
$$
h=\ints (\Lambda^{(3)} i \sqrt{2} \partial X^{26}
+ \Lambda^{(+)} \exp({i}{\sqrt 2}X^{26}_L)
+{\Lambda^{(-)}} \exp({-i}{\sqrt 2}X^{26}_L)),
\numbereq
$$
where the functions ${\Lambda^{(\alpha)}}$ are functions of
$X^\mu$ only.  (There are similar currents and transformations
for $SU(2)_R$.)

Away from the critical radius, the current $J^{(3)}$ deforms,
but the $SU(2)_L$ symmetry algebra continues to hold.
For $R \ne R_{cr}$, however, the currents $J^{(\pm)}\s$ are not
of dimension $(1,0)$ with respect to the deformed stress energy
tensor.  This implies that the currents are not conserved, and
the spacetime symmetry is spontaneously broken.

In this section we will study the Higgs effect by implementing
an arbitrary $SU(2)_L$ gauge transformation for $R \ne R_{cr}$.
We shall start at the critical radius, and turn on a constant
value for the field $\phi^{(33)}$.  We shall see that this
defines a new stress tensor $T'\s$ which is equivalent to the
stress tensor $T_R\s$ at a radius $R = R_{cr} + \delta R = R_{cr}
(1 - \langle \phi^{(33)} \rangle)$.  We will then turn on fields
$\phi^{(\pm3)}(X)$ and $A^{(\pm)}_\mu(X)$.  This defines a new stress
tensor $T''\s$ which describes infinitesimal fluctuations of the
would-be Goldstone bosons $\phi^{(\pm3)}(X)$ and the broken-symmetry
gauge fields $A^{(\pm)}_\mu(X)$ about the string vacuum at radius $R$.
Once we have the stress tensor $T''\s$, we will compute an arbitrary
broken-symmetry gauge transformation.  We will see that the would-be
Goldstone fields transform by a shift.  This will permit us to pass
to unitary gauge and identify the physical fields.

Therefore let us start at the radius $R_{cr}$ and turn on a constant
value for the field $\phi^{(33)}$.  To first order, the stress
tensor is just
$$
T'\s = -{1\over 2} \eta_{\mu\nu} \partial X^\mu \partial X^\nu -
{1\over 2} \partial X^{26} \partial X^{26}
- \langle \phi^{(33)}\rangle \partial X^{26}
\overline{\partial} X^{26} + \ldots .
\numbereq\name{\phivev}
$$
Conformal invariance is satisfied because $\langle \phi^{(33)}
\rangle$ is constant.  Comparing (\phivev) with (\eqsciopp), we
see that $\delta R$ can be identified with $- \langle \phi^{(33)}
\rangle R_{cr}$.

Let us now deform this stress tensor by turning on the fields
$\phi^{(\pm3)}(X)$ and $A^{(\pm)}_\mu$(X), following the
techniques of sect.~2.  Therefore we add to $T'\s$
a deformation of the form
$$
\eqalign{
\delta T' =&-\phi^{(+3)}(X)\hat{\overline{\partial}}X^{26}
\exp({i}{\sqrt 2}X^{26}_L)- \phi^{(-3)}(X)\hat{\overline{\partial}}
X^{26} \exp({-i}{\sqrt 2}X^{26}_L)\cr
&-A^{(+)}_\mu(X) \overline{\partial}X^\mu
\exp({i}{\sqrt 2}X^{26}_L)- A^{(-)}_\mu(X) \overline{\partial}X^\mu
\exp({- i}{\sqrt 2}X^{26}_L),\cr}
\numbereq
$$
where the hatted derivatives are given by (\eqsmz), with $\delta R
= - \langle \phi^{(33)} \rangle R_{cr}$.
This deformation is conformal if the functions $\phi^{(\pm3)}(X)$
and $A^{(\pm)}_\mu(X)$ obey the conditions (\eqaxoin).  The resulting
stress tensor $T''\s = T'\s + \delta T'$ describes a consistent
string background with excitations of the broken-symmetry gauge
bosons and the would-be Goldstone bosons around a vacuum with
arbitrary radius $R$ -- that is, a non-zero vacuum expectation value
for the spacetime field $\phi^{(33)}(X)$.

We now perform a local gauge transformation generated by
$$
h=\ints (\Lambda^{(+)}\exp({i}{\sqrt 2}X^{26}_L)
+\Lambda^{(-)}\exp({-i}{\sqrt 2}X^{26}_L)).
\numbereq\name{\eqaut}
$$
The operator $h$ contains the $SU(2)_L$ gauge transformations
which do not respect the string vacuum.  As in sect.~2, we compute
the commutator $i [h,T''(\sigma)]$ and find
$$
\eqalign{
T''\s + i [h_1,T''(\sigma)] =&
-(\phi^{(+3)}(X)+ \delta R \Lambda^{(+)}(X))\hat{\overline{\partial}}X^{26}
\exp({i}{\sqrt 2}X^{26}_L)\cr
& -(\phi^{(-3)}(X)+ \delta R \Lambda^{(-)}(X))\hat{\overline{\partial}}
X^{26} \exp({-i}{\sqrt 2}X^{26}_L)\cr
& -(A^{(+)}_\mu(X) + \partial_\mu \Lambda^{(+)}(X))
\overline{\partial}X^\mu \exp({ i}{\sqrt 2}X^{26}_L)\cr
& -(A^{(-)}_\mu(X) + \partial_\mu \Lambda^{(-)}(X))
\overline{\partial}X^\mu \exp({- i}{\sqrt 2}X^{26}_L).\cr}
\numbereq
$$
Defining
$$
\eqalign{
\phi'^{(\pm3)}(X) &= \phi^{(\pm3)}(X) + \delta R \Lambda^{(\pm)}(X)\cr
A'^{(\pm)}_\mu(X) &= A^{(\pm)}_\mu(X) + \partial_\mu \Lambda^{(\pm)}
(X), \cr}
\numbereq
$$
we see we can write $T'''\s = T''\s + i [h_1,T''(\sigma)]$ as
$$
\eqalign{
T'''\s =& T'\s
- \phi'^{(+3)}(X) \hat{\overline{\partial}}X^{26} \exp({i}{\sqrt 2}X^{26}_L)
- \phi'^{(-3)}(X) \hat{\overline{\partial}}X^{26} \exp({-i}{\sqrt
2}X^{26}_L)\cr
& - A'^{(+)}_\mu(X) \overline{\partial}X^\mu \exp({i}{\sqrt 2}X^{26}_L)
- A'^{(-)}_\mu(X) \overline{\partial}X^\mu \exp({- i}{\sqrt 2}X^{26}_L),
\cr}
\numbereq
$$
where the conformal condition (\eqaxoin) is the gauge-covariant
equation of motion,
$$
\Box A'^{(\pm)}_\mu(X)-\partial_\mu\partial^\nu
A'^{(\pm)}_\nu(X)+{(\delta R)^2\over 2}A'^{(\pm)}_\mu(X)
= \left({\delta R\over R_{cr}^2}-{(\delta R)^2\over 2R_{cr}^3}\right)
\partial_\mu \phi'^{(\pm3)}(X).
\numbereq\name{\eqwsiora}
$$
{}From this we see that the broken-symmetry gauge bosons and the
would-be Goldstone bosons obey coupled equations of motion.

The stress energy tensor $T'''\s$ describes a string vacuum that
is physically equivalent to that of $T''\s$.  Note that under
the automorphism (\eqaut), the would-be Goldstone bosons
transform by a shift,
$$
\phi^{(\pm3)}(X) \mapsto \phi^{(\pm3)}(X)+\delta R \Lambda^{(\pm)}(X).
\numbereq\name{\eqrcosn}
$$
This confirms that the would-be Goldstone bosons are gauge
artifacts, and that they can be transformed away by a suitable gauge
transformation.

To exhibit the Higgs effect explicitly, let us
choose the transformation parameters to eliminate the would-be
Goldstone bosons from the spectrum,
$$
\delta R \Lambda^{(\pm)}(X)=-\phi^{(\pm3)}(X)
$$
In this unitary gauge, the stress tensor reduces to
$$
T'''\s \rightarrow T'\s - A'^{(+)}_\mu(X) \overline{\partial}X^\mu
\exp({ i}{\sqrt 2}X^{26}_L)
- A'^{(-)}_\mu(X) \overline{\partial}X^\mu \exp({- i}{\sqrt 2}
X^{26}_L).
\numbereq
$$
The conformal condition (\eqwsiora) reduces to a set of massive
equations of motion for the vectors $A'^{(\pm)}_\mu(X)$,
$$
\Box A'^{(\pm)}_\mu(X)-\partial_\mu\partial^\nu
A'^{(\pm)}_\nu(X)
+{(\delta R)^2\over 2}A'^{(\pm)}_\mu(X)
=0 .
\numbereq
$$
The would-be Goldstone bosons have become the longitudinal components
of the massive gauge bosons.

It is straightforward to verify that the number of physical
states does not change as the radius is varied.  Indeed, at the
critical point, the spectrum includes the massless gauge bosons
$A^{(\pm)}_\mu$, with twenty-three polarizations each, as well as
the real scalar fields $\phi^{(\pm3)}$.
Away from the critical point, the scalars are gone,
but the gauge bosons are massive, with twenty-four polarizations
each, so the total number of degrees of freedom remains the
same.

\newsection Conclusions.

In this paper we illustrated the Higgs mechanism in string
theory.  We considered the simple example of string propagation on
$M^{24,1} \times S^1$, but our procedure may be readily generalized
to other string backgrounds.
We started with the operator algebra $\cal A$, at the critical radius
of $S^1$, where the symmetry algebra is $SU(2)_L \times SU(2)_R$.
Using a fixed basis of operators, we constructed the stress tensor
and vertex operators for the gauge and scalar bosons, as well as
the generators of the $SU(2)_L \times SU(2)_R$ symmetry algebra.

We then deformed the conformal field theory by varying the radius of
the circle away from its critical value.  We studied the change in
the vertex operators along the deformation class.  We found the
$SU(2)_L \times SU(2)_R$ gauge symmetry to be spontaneously broken
to $U(1)_L \times U(1)_R$, and the corresponding world-sheet currents
to be no longer conserved.

In the final section of the paper, we derived the gauge-covariant
equation of motion for the broken-symmetry gauge bosons and the
would-be Goldstone bosons.  We eliminated the scalars
from the spectrum by performing a suitable gauge transformation.  In
this unitary gauge, the would-be Goldstone bosons became
the longitudinal components of the massive gauge bosons.

\immediate\closeout1
\bigbreak\bigskip

\line{\bf Acknowledgements. \hfil}
We would like to thank M.~Evans and J.~Liu for useful discussions.
This work was supported by the U.S.~National Science Foundation,
grant NSF-PHY-9404057, and the U.S.~Department of Energy
Contract Number DE-FG02-91ER40651-TASKB

\nobreak\medskip

\line{\bf References.\hfil}
\nobreak\medskip\vskip\parskip

\input refs

\end